# Self-Interacting Electron as a Nonlinear Dynamical System

Vladimir A. Manasson
(WaveBand Division of Sierra Nevada Corporation, Irvine, California
e-mail: vladimir.manasson@sncorp.com)

**Abstract**

We have proposed a simple one-dimensional model of internal particle dynamics. The model is based on the assumption that self-interaction can be represented by a nonlinear feedback and described by a quadratic recurrent map. Charge plays the role of a generalized dynamical variable and a feedback coupling parameter. The model suggests that charge and action quantization stem from the system's dissipative quality and from a hierarchy of supercycle orbits located between period-doubling bifurcations on the Feigenbaum tree. Among the numerical results, we have discovered a link between the quantum of action and the elementary charge. We also found that the fine structure constant can with a good accuracy be expressed exclusively through mathematical constants, including the Feigenbaum delta. We have introduced dimensionless numbers that describe the relative role of the internal particle dynamics when both internal and external dynamics are taken into consideration. We have found these numbers to be close to the electron, proton, and neutron g-factors known from the experiment.

## 1. Introduction

Let us consider a negatively charged vacuum fluctuation. It polarizes the surrounding vacuum and creates a positively charged halo. The halo lowers the local electric potential and the original fluctuation becomes more confined and dense. This in turn affects the halo. This positive feedback competes with a negative feedback caused by charge diffusion and self-repellence. The symmetric halo becomes unstable and brakes up into separate positively charged fragments. Those in turn create secondary negatively charged halos around themselves. The process repeats itself infinitely at different scales. In reality, this *fractal* picture is even more complex because the process is a dynamic one and also involves currents and magnetic fields. The initial fluctuation may die, or else engender more or less stable spatially extended dynamic patterns. We conjecture that elementary particles are manifestations of stable spatio-temporal patterns in the dynamic vacuum.

To make this self-organizing scenario live, vacuum must also be a dissipative medium. One testimony to vacuum's dissipative quality is the cooling relict background radiation. In nonlinear dissipative media, fluctuations can grow into *repeatable* spatio-temporal structures [1-4]. Examples of such structures are: vortexes, domain walls, various sorts of waves, Bénard cells, linear and point defects, and others. Those patterns have been found in various media, such as moving fluids, the atmosphere, chemical reactions, gaseous and solid-state plasmas, laser cavities, electric circuits, and cellular automata.

Electron self-interaction based on the picture described above is far too complex for full-scale modeling. However, some degree of understanding and even some quantitative results may be obtained from studying simplified models that reflect the most essential features of the system. In our case, this hope is supported by the universal nature of the relevant models known in nonlinear dynamics [5-7] and their surprising effectiveness in describing complex behavior of real systems [8-16]. One reason for this success is, again, the dissipative nature of the systems. This dissipative quality can effectively lower the dimensionality of the phase space of the respective dynamic system. Dissipative systems do not preserve the phase-space volume occupied by the system's trajectories. In the course of its general contraction, the phase volume varies in different directions at different rates. In some directions it shrinks, in others it even grows and folds onto itself. Ultimately, it converges to an almost *one-dimensional* line [7,17].

## 2. Model

In this paper we propose a model of charge quantization. The model assumes that the electron can be represented as a dissipative dynamic system and that delayed nonlinear feedback is an essential feature of this system. We also assume that the internal dynamics of the self-interacting electron can be effectively described by trajectories spiraling toward closed orbits (attractors) in a low-dimensional phase space.

Being near a closed orbit, the system behaves like a conservative one. By proper selection of the phase space, the system can be described in terms of generalized action-angle variables, where the angle $\varphi$ is a cyclic or ignorable coordinate, and the conjugated generalized momentum $J$ is a conserved parameter characterizing the respective orbit [18]. Since $\varphi$ is dimensionless $J$ should have the dimensions of angular momentum or action. However, the electron is an entity that is preferably characterized by charge. Charge quantization is the major objective of our exploration. Therefore, we will introduce a new variable. Like action, it too is a conserved variable. We will call it generalized charge, $q$. It has the same dimensions as conventional electric charge and will play in the internal electron dynamics a role similar to the one the conventional charge plays in the external electron dynamics, namely it will reflect the coupling strength between the particle and the internal electromagnetic field, i.e. the electromagnetic field that is generated and absorbed within the particle.

Particle internal dynamics is relativistic. Traditional dynamic variables, such as energy and momentum, are components of the momentum four-vector and depend on the selected coordinate system. In the course of the phase-volume metamorphosis that accompanies transition to the low-dimensional dynamics, spatial coordinates and time merge in a complex way, much more complex than simple rotations in a four-dimensional space. In this transition, we loose information about four-vector components and their interrelations. In contrast, charge is a Lorentz scalar and thus is much more appropriate to be used in our low-dimensional model.

To link *J* and *q,* let us keep in mind that electron self-interaction is a process that involves the electromagnetic field created by the particle itself, as well as the reaction of this field back onto the electron. Action for a charge $q$ interacting with an electromagnetic field $A_i$ is given by the integral $S = -(q/c)\int_a^b A_i dx^i$ [19], where *c* is the velocity of light, $A_i$ is the four-potential, $x_i$ are space-time coordinates and the integration is taken along a world line. In a system representing exclusively a solitary self-interacting electron (there is no external field), the electron is the only object responsible for the origin of $A_i$. Four-potential is a four-vector whose components are proportional to the electron charge either directly or via the currents generated by the charge. In our low-dimensional model, we need to replace the four-potential with a scalar entity playing the same role. We don't know the explicit form of this scalar but we assume that like $A_i$ it should be proportional to *q*. Thus in our model, the action *S* and the generalized momentum *J* will both be proportional to $q^2$. We will define *J* as

$$J = \eta q^2, \tag{1}$$

where a conversion constant $\eta = \sqrt{\mu_0/\varepsilon_0}$ is introduced to "match" the dimensions of *J* with those of $q^2$ ($\eta = 376.73...\,\Omega$ is the vacuum impedance, $\varepsilon_0$ is vacuum permittivity, and $\mu_0$ is vacuum permeability). We will define the action accumulated in one cycle of "motion" along the closed orbit as

$$S = -\int_0^{2\pi} J d\varphi = -2\pi\eta q^2. \tag{2}$$

To further simplify the description, we will replace each continuous trajectory with a stroboscopic set of points $x_i$ that belong to the trajectory (one point per loop) and have been selected using a procedure called Poincaré section [17]. As a result, we will obtain a one-dimensional recurrent map

$$x_k = F_q(x_{k-1}), \tag{3}$$

where *F(x)* is a recursion function parameterized by *q*. To specify the function *F(x)* we shall use the analogy between our system and other nonlinear systems with a complex feedback described by the logistic map [17]:

$$x_k = ax_{k-1} - ax_{k-1}^2. \tag{4}$$

The logistic map comprises positive (first term on the right) and negative (second term on the right) delayed feedback. The quadratic term implies nonlinearity. The strength of the feedback is determined by the parameter *a*. In a self-interacting electron, the delayed feedback is implemented via the electron/electromagnetic field interaction and the feedback is proportional to the coupling strength between the electron and the field, which is represented by the charge $q$. We will use the parallel between the parameters $q$ and $a$ assuming that as a result of varying the parameter $q$, the map (3) behaves the same way as the logistic map behaves as a result of varying parameter $a$. This similarity in behavior prevails for a wide class of functions $F(x)$. It is only important that $F(x)$ be unimodal (i.e. possess a single extremum within the entire interval of its definition) and smooth [17]. The similarity, in some important aspects, is even *quantitative* if the function $F(x)$ is quadratic near its extremum. The latter restriction is not very limiting, as many smooth functions can be expanded near the extremum into a Taylor series, with the quadratic term being the dominant one.

Despite its apparent simplicity, the map (3) encompasses a rich and complex dynamics [5-7, 17]. For small values of the parameter $q$, there is an interval where the map converges to fixed points, indicating the existence of closed orbits. For another $q$-interval, the map (3) does not converge but its second iterated map $x_k = F_q(F_q(x_{k-1}))$ does, implying the existence of period-2 closed orbits. With increasing values of $q$, stability shifts from period-2 orbits to period-4 orbits, from period-4 orbits to period-8 orbits and so on. The dynamics can be depicted as an infinite cascade of period-doubling bifurcations, shown in Fig.1a in a bifurcation diagram. This fractal represents a tree, each branch of which mimics the entire tree but at a different scale. The number of branches at each level $q$ corresponds to the number of loops in period-$2^n$ closed orbits. The scaling factor along the $q$-axis is a number quickly converging to $\delta$, a universal constant called the Feigenbaum delta. For all unimodal iterated maps with function $F(x)$ behaving quadratically near its extremum, $\delta = 4.669202\ldots$[5, 17].

Between bifurcations the system is stable, and trajectories converge to the closed orbits parameterized by $q$. However, the convergence rate is not the same for different $q$s. The convergence rate depends on the derivative of the respective iterated function, and the convergence rate is maximum for trajectories whose fixed point is located near the function's extremum, $x_{ext}$ [17, 20]. At $x_{ext}$ all derivatives

$$\frac{d}{dx}\left(F\left(F\left(\ldots\left(F\left(x_{ext}\right)\right)\right)\right)\right) = 0. \tag{5}$$

The corresponding limit cycles (they are called supercycles [17]) occur for specific values of the parameter $q$ (we denote them as $q_i$). The supercycle parameter values $q_i$ can be found on the bifurcation tree where the branches are crossing the line $x_{ext}$ (shown by circles in the Fig.1 diagram).

We can conditionally divide each converging trajectory into two portions, one being a transient spiral, the other one being an attractor, an almost closed orbit. For the selected trajectory, the higher the convergence rate the shorter is the transient time, the longer is the time the system spends near the attractor, and the higher is the probability of finding the system near the attractor. Vacuum fluctuations or other external noise are responsible for kicking the electron from one phase trajectory to another. Due to the different convergence rates, supercycles are the most "attractive" attractors, and thus the probability of finding the system near a supercycle is higher than the probability of finding it in other portions of the phase space. The existence of specific most probable discrete orbits means their quantization, as well as the quantization of the parameters $q$ and $S$ associated with those orbits, i.e. *quantization of charge and action.*

We can generalize the definition of the action (2) for period-$2^k$ supercycles as:

$$S_k = -\int_0^{2^k \pi} J_k \, d\varphi = -2^k \pi \eta q_k^2 . \tag{6}$$

The supercycle parameters $q_k$ obey the same scaling law as the entire bifurcation tree, with the same asymptotic behavior converging to the Feigenbaum delta [17]:

$$\lim_{k \to \infty} (q_{k-1}/q_k) = \delta . \tag{7}$$

From (6) and (7), the ratio between two actions accumulated along two neighboring supercycle orbits converges as

$$\lim_{k \to \infty}(S_{k-1}/S_k) = \lim_{k \to \infty}(q_{k-1}/q_k)^2 / 2 = \delta^2/2 . \tag{8}$$

The convergence rate in (7) is usually high [5, 17] and even for relatively small numbers $k$ we can, to a good approximation, assume the exact equalities:

$$q_{k-1}/q_k = \delta \tag{9}$$

and

$$S_{k-1}/S_k = \delta^2/2 . \tag{10}$$

Based on the proposed model, the equilibrium states of the self-interacting electron can be represented by a set of discrete orbits, or a hierarchy of levels with quantized charges $q_i$ and actions $S_i$. Formulas (9) and (10) provide relations between charges and actions at different levels. The deepest level corresponds to the electron dynamics confined to single-loop closed trajectories. With increasing $i$, the trajectory becomes more complex and is spread over a larger phase volume. Above the critical level $S_\infty$ which indicates transition to chaos (it is called the Feigenbaum point), the trajectories fill the phase volume continuously and the discrete pattern disappears. When the system locates at a shallow level, an external noise, such as background radiation, can effectively disrupt the picture by kicking the system from an attractor to the spiraling part of the same trajectory or to another trajectory.

As already mentioned, the bifurcation tree is a fractal, each branch mimicking the entire tree. If we are interested only in the dynamics involving level $k$ and above, we can replace the original tree with a truncated one, such that the level $k$ of the original tree plays the role of the principal level in the new tree. In the truncated tree we should assign new numbers to the supercycles $(m + k - 1) \to m$ and use renormalized values for charge $\tilde{q}_m$ and action $\tilde{S}_m$ which are given by

$$\tilde{q}_m = q_{m+k-1} \tag{11}$$

and

$$\tilde{S}_m = -2^m \pi \eta \tilde{q}_m^2. \tag{12}$$

### 3. Numerical Estimates

We have built a model for internal electron dynamics that describes the particle's interaction with itself via the electromagnetic field created and absorbed by the particle itself. However, the internal particle world is inaccessible to direct measurements. To make any numerical estimates, we need to relate the internal particle dynamics to its external dynamics and use the parameters that can be obtained from experiment. Two universal physical constants, which are counterparts of the dynamical variables selected for the proposed model, characterize electromagnetic interactions: the elementary charge $e = 1.60217... \times 10^{-19}\, C$ and the action accumulated during one cycle of photon oscillation which is expressed by the Planck's constant $h = 6.626068.. \cdot 10^{-34}\, J \cdot s$.

To link the internal dynamics to the external dynamics, we will assume that our model is applicable to the description of the external dynamics and that the coupling between the particle and the external field is given by $q_{external}$, which is also a conserved dynamical variable but this time used for external interactions. In the case of the electron, $q_{external} = -e$. When an electron interacts with an external electromagnetic field it is impossible to distinguish the electromagnetic field created by the particle from the external field. Therefore, we conjecture that the internal electron dynamics merges with its external dynamics, and that both dynamics belong to the same level, say level $n$.

In the following discussion we will explore the internal dynamics only at two levels, the level $n$ and one level below. It will be convenient to truncate the bifurcation tree below the level $n-1$ and renormalize the charge and action (see Fig.1b). We will assign to the level $n-1$ and level $n$ new numbers, 1 and 2, respectively.

We assume that for the level 2, the renormalized charge is

$$q_{\text{int }ernal} = q_n = \tilde{q}_2 = q_{external} = -e. \tag{13}$$

Then, the action at this level is

$$\left|\tilde{S}_2\right| = 2\pi\eta e^2 \approx 0.1h << h. \tag{14}$$

Using (10) and (13) we can also estimate the values of charge and action at level 1:

$$\tilde{q}_1 = -\delta e \tag{15}$$

and

$$\left|\tilde{S}_1\right| = \left|\delta^2 \tilde{S}_2/2\right| = \pi\eta\delta^2 e^2. \tag{16}$$

Numerical calculation shows that the term on the right hand side is

$$\pi\eta\delta^2 e^2 = 6.623482..\times 10^{-34}\, J\cdot s, \tag{17}$$

which is surprisingly close to the Planck's constant. The two differ only by

$$\left(h - \pi\eta\delta^2 e^2\right)/h \approx 4\cdot 10^{-4}. \tag{18}$$

With the same accuracy, the reduced Planck's constant $\hbar = h/2\pi$ can be expressed as

$$\hbar = \eta\delta^2 e^2/2, \tag{19}$$

and the fine structure constant $\alpha = e^2/4\pi\varepsilon_0 c\hbar$ can be expressed as

$$2\pi\alpha = \delta^{-2}, \tag{20}$$

i.e. in terms of only mathematical constants: $\pi$ and $\delta$.

Let us now explore the situations when the internal dynamics belongs to different levels, while the external dynamics still belongs to the level 2 (for all electromagnetic interactions, experiment gives the same absolute value for the coupling constant, $e$). To compare different situations, it will be convenient to introduce a new dynamical variable, the "total charge" $Q^{\pm} = q_{external} \pm q_{internal}$. The total charge, too, is a conserved dynamical variable. It describes the combined external/internal dynamics. Selection of "+" or "–" in front of the second term on the right depends on the "sense" of rotation on the internal orbit. The total charge can be expressed in units of the elementary charge. The corresponding dimensionless number will be expressed by a parameter $g_i^{\pm}$

$$g_i^{\pm} = \frac{Q^{\pm}}{e}, \tag{21}$$

where the number $i$ indicates the level to which the internal dynamics belongs.

Let us consider several cases:

a) there is no internal dynamics ($i = \infty$, $q_{internal} = 0$); $q_{external} = \pm e$. From Eq. (21) we find that

$$g_\infty = \pm 1 . \qquad (22)$$

b) the internal dynamics belongs to the level 2, and $q_{internal} = q_{external} = -e$. From Eq. (21) we find that

$$g_2^+ = -2 . \qquad (23)$$

c) the internal dynamics belongs to the level 1 and $q_{internal} = +\delta e$; $q_{external} = +e$. From Eq. (21) we find that

$$g_1^+ = \frac{e + \delta e}{e} = 1 + \delta \approx 5.669... \qquad (24)$$

and

$$g_1^- = \frac{e - \delta e}{e} = 1 - \delta \approx -3.669... . \qquad (25)$$

What is special about these g-numbers? According to our model, they express the role of the internal dynamics. For particles without internal dynamics $|g| = 1$, for particles with internal dynamics it is different from 1. We used the letter "g" not by accident. In particle physics, the role of the particle's internal dynamics is expressed by its g-factor.

For example, for a classical pointlike particle, i.e. a particle that does not possess an internal structure, the g-factor is equal to $\pm 1$ (the selection of the sign depends on the charge polarity). For electron, proton, and neutron, the measured values of g-factors are -2.002…, +5.5856…, -3.8260…. What is remarkable, that the g-numbers from Equations (23), (24), and (25) differ from g-factors for these particles only by 0.1%, 1.5%, and 4%, respectively. This is another surprise of the proposed model.

We can use these results to make a sort of particle classification. The particle taxonomy is shown in Table 1.

Table 1. Particle taxonomy according to the proposed model.

| Particle | Internal Dynamics | | | External Dynamics | | | g-number |
|---|---|---|---|---|---|---|---|
| | Level | Coupling constant | Action | Level | Coupling constant | Action | |
| *Pointlike* | ∞ | 0 | 0 | 2 | $\pm e$ | $2\pi\eta e^2$ | $\pm 1$ |
| $e$ | 2 | $-e$ | $2\pi\eta e^2$ | 2 | $-e$ | $2\pi\eta e^2$ | $-2$ |
| $p$ | 1 | $\delta e$ | $\pi\eta\delta^2 e^2 = h$ | 2 | $e$ | $2\pi\eta e^2$ | $1+\delta$ |
| $n$ | 1 | $\delta e$ | $\pi\eta\delta^2 e^2 = h$ | 2 | $e$ | $2\pi\eta e^2$ | $1-\delta$ |
| $\gamma$ | 1 | $\pm \delta e$ | $\pi\eta\delta^2 e^2 = h$ | ∞ | 0 | 0 | 0 |

We have added one more particle to the list of actors. It is the photon. Photon interacts with particles but does not interact with the external electromagnetic field. According to our classification this means that photon does not possess any external dynamics, i.e. we can assign to it an external dynamics level $i = \infty$. However photon does possess internal dynamics and this dynamics is described by the action $h$. Therefore we placed the photon at the level 1.

The Feigenbaum bifurcation tree has one more interesting feature. Each bifurcation point represents a phase transition, namely period doubling. In accordance with our model, the number of cycles per full period at the level 1 is as twice as that at the level 2. External electron-photon interaction involves dynamics at different levels: electron is at the level 2 and photon is at the level 1. Assuming that during the interaction their motion is synchronized, this means that during one cycle of the photon's oscillation, the electron completes only half of a full cycle of rotation. This electron period-doubling can be considered as manifestation of the SU(2) symmetry that describes spin-1/2 particles. The same is true for nucleons whose external dynamics also belongs to the level 2.

## 4. Conclusions

Nonlinear dynamics and chaos theory provide new approaches to the old problems. We have proposed a simple model of charge quantization based on the assumption that the electron's internal dynamics obeys the Feigenbaum period-doubling scenario. Despite the fact that our model is qualitative and rather heuristic, we have obtained several sensible numerical results. The most unexpected suggests a link between two fundamental constants, the elementary charge $e$ and the Planck's quantum of action $h$. Another surprising result is the closeness of the g-numbers characterizing the role of internal dynamics to the particle g-factors.

## Acknowledgements

The author is grateful to Ryszard Gajewski, Vladimir Litvinov, and Lev Sadovnik for critical reading of the manuscript, and Victor Dugaev for encouraging.

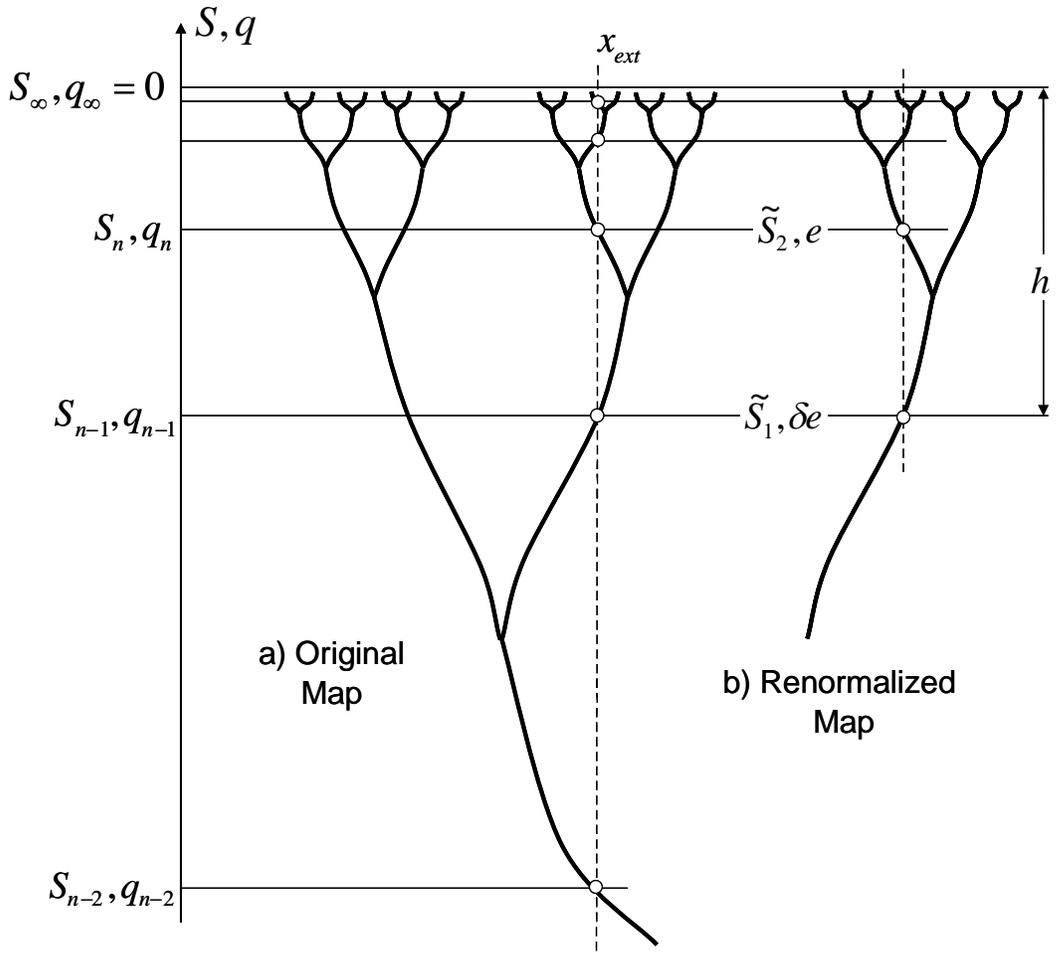